\documentclass[10pt, conference, letter]{IEEEtran}
\IEEEoverridecommandlockouts

\usepackage{algorithmic}
\usepackage{amsmath,amssymb,amsfonts}
\usepackage{booktabs}
\usepackage{cite}
\usepackage{enumitem}
\usepackage{fancyhdr}
\usepackage{glossaries}
\usepackage{graphicx}
\usepackage[utf8]{inputenc}
\usepackage{multirow}
\ifCLASSOPTIONcompsoc
    \usepackage[caption=false,font=normalsize,labelfont=sf,textfont=sf]{subfig}
\else
    \usepackage[caption=false,font=footnotesize]{subfig}
\fi
\usepackage{textcomp}
\usepackage{todonotes}
\usepackage{xcolor}
\graphicspath{{figs/}}

\def\BibTeX{{\rm B\kern-.05em{\sc i\kern-.025em b}\kern-.08em
    T\kern-.1667em\lower.7ex\hbox{E}\kern-.125emX}}

\definecolor{darkred}{rgb}{0.5,0,0}
\newcommand{\toreview}[0] {
  \color{black}}

\newacronym{ALU}   {ALU}      {Arithmetic-Logic Unit}
\newacronym{BB}    {BB}       {Body Biasing}
\newacronym{BT}    {BT}       {Binary Translation}
\newacronym{CCA}   {CCA}      {Configurable Compute Accelerator}
\newacronym{CMP}   {CMP}      {Chip Multi-Processor}
\newacronym{CGRA}  {CGRA}     {Coarse-Grained Reconfigurable Array}
\newacronym{DBT}   {DBT}      {Dynamic Binary Translation}
\newacronym{DFG}   {DFG}      {Data-Flow Graph}
\newacronym{DIM}   {DIM}      {Dynamic Instruction Merging}
\newacronym{DORA}  {DORA}     {Dynamic Optimizer for Reconfigurable Architectures}
\newacronym{DynaSpAM}{DynaSpAM}{Dynamic Spatial Architecture Mapping}
\newacronym{DVFS}  {DVFS}     {Dynamic Voltage and Frequency Scaling}
\newacronym{EDP}   {EDP}      {Energy-Delay Product}
\newacronym{FPGA}  {FPGA}     {Field-Programmable Gate Array}
\newacronym{FU}    {FU}       {Functional Unit}
\newacronym{f}     {$f$}      {Operating Frequency}
\newacronym{GPP}   {GPP}      {General-Purpose Processor}
\newacronym{GPU}   {GPU}      {Graphics Processing Unit}
\newacronym{HCI}   {HCI}      {Hot-Carrier Injection}
\newacronym{HDL}   {HDL}      {Hardware Description Language}
\newacronym{ILP}   {ILP}      {Instruction-Level Parallelism}
\newacronym{ISA}   {ISA}      {Instruction-Set Architecture}
\newacronym{IoT}   {IoT}      {Internet-of-Things}
\newacronym{NTC}   {NTC}      {Near-Threshold Computing}
\newacronym{NTV}   {NTV}      {Near-Threshold Voltage}
\newacronym{NBTI}  {NBTI}     {Negative-Bias Temperature Instability}
\newacronym{PVT}   {PVT}      {Process, Voltage, Temperature}
\newacronym{OoO}   {OoO}      {Out-Of-Order}
\newacronym{OS}    {OS}       {Operating System}
\newacronym{PC}    {PC}       {Program Counter}
\newacronym{PE}    {PE}       {Processing Element}
\newacronym{QoS}   {QoS}      {Quality-of-Service}
\newacronym{RTL}   {RTL}      {Register-Transfer Level}
\newacronym{ROB}   {ROB}      {Re-Order Buffer}
\newacronym{SoC}   {SoC}      {System-on-Chip}
\newacronym[firstplural=Static Random-Access Memories (SRAMs)]{SRAM}  {SRAM}     {Static Random-Access Memory}
\newacronym{STC}   {STC}      {Super-Threshold Computing}
\newacronym{STV}   {STV}      {Super-Threshold Voltage}
\newacronym{TLP}   {TLP}      {Thread-Level Parallelism}
\newacronym{TransRec}{TransRec}{Transparent Reconfigurable}
\newacronym{Vdd}   {$V_{dd}$} {Operating Voltage}
\newacronym{Vt}    {$V_{t}$}  {Threshold Voltage}

\begin{document}

\bstctlcite{IEEEexample:BSTcontrol}

\title{Proactive Aging Mitigation in CGRAs through Utilization-Aware Allocation
    \vspace{-2mm}}

\author{
    \IEEEauthorblockN{%
    Marcelo Brandalero$^{\ast}$, %
    Bernardo Neuhaus Lignati$^{\dagger}$,
    Antonio Carlos Schneider Beck$^{\dagger}$,\\
    Muhammad Shafique$^{\ddagger}$, %
    Michael Hübner$^{\ast}$}
    \IEEEauthorblockA{
    \textit{$^{\ast}$ Chair of Computer Engineering, Brandenburg University of Technology (B-TU).} Cottbus, Germany. \\
    \textit{$^{\dagger}$ Institute of Informatics, Universidade Federal do Rio Grande do Sul (UFRGS).} Porto Alegre, Brazil. \\
    \textit{$^{\ddagger}$ Institute of Computer Engineering, Vienna University of Technology (TU Wien).} Vienna, Austria. \\}
    \vspace{-6mm}
}

\maketitle

\thispagestyle{fancy}
\fancyhf{} 
\fancyhead[C]{To appear at the 57th Design Automation Conference (DAC), July 2020, San Francisco, CA, USA.}
\fancyfoot[C]{}

\begin{abstract}
Resource balancing has been effectively used to mitigate the long-term aging effects of Negative Bias Temperature Instability (NBTI) in multi-core and Graphics Processing Unit (GPU) architectures.
In this work, we investigate this strategy in Coarse-Grained Reconfigurable Arrays (CGRAs) with a novel application-to-CGRA allocation approach.
By introducing important extensions to the reconfiguration logic and the datapath, we enable the dynamic movement of configurations throughout the fabric and allow overutilized Functional Units (FUs) to recover from stress-induced NBTI aging.
Implementing the approach in a resource-constrained state-of-the-art CGRA reveals $2.2\times$ lifetime improvement with negligible performance overheads and less than $10\%$ increase in area.
\end{abstract}


\glsresetall

\section{Introduction}
Hardware aging has emerged as a critical reliability threat that can lead to degraded performance and early-stage system failure \cite{Henkel2011}.
One of the main phenomena leading to increased \gls{Vt}, circuit delays, and device wear-out is \gls{NBTI}, caused by the massive stress induced on the PMOS transistors \cite{Bernstein2006, Rahimi2013a}.
Partial recovery from this aging threat is observed when the circuit is power-gated or remains idle for a given time.
Utilization-aware resource balancing can thus be leveraged to alleviate the long-term aging effects of \gls{NBTI}.
Recent approaches to aging mitigation exploit the architectural regularity in \glspl{CMP} and \glspl{GPU} with scheduling algorithms that balance the utilization of the processing resources (i.e., cores) and reduce the accumulated stress, thereby slowing down the wear-out \cite{Gnad2015, Paterna2013, Muck2017, Rahimi2013a, Chen2014a, Lee2017, Feng2010, Bolchini2016}.
However, within the scope of small low-power \glspl{CMP} or even single-core embedded systems, the absence of a regular structure hinders the application of similar strategies. 

Reconfigurable architectures, in particular, \gls{CGRA} systems, offer a promising solution in that direction.
Besides being dynamically customizable to implement different application datapaths that improve performance and energy efficiency, \glspl{CGRA} also offers a regular structure that can be leveraged by utilization-balancing strategies for aging mitigation.
While previous works have addressed energy-efficient design and mapping strategies for \glspl{CGRA} \cite{Hamzeh2012, Chen2014b, Dave2018, Clark2004, Beck2014, Liu2015, Watkins2016}, aging mitigation in such architectures has received significantly less attention, with the proposed strategies being either oblivious to the utilization \cite{Afzali-Kusha2018} or requiring changes in the compiled code \cite{JiangyuanGu2017}.
To address these limitations in \glspl{CGRA}, in this paper, we propose an automatic approach for aging mitigation through utilization-aware resource balancing.

\begin{figure}[t]
    \centering
    \includegraphics[width=0.95\linewidth]{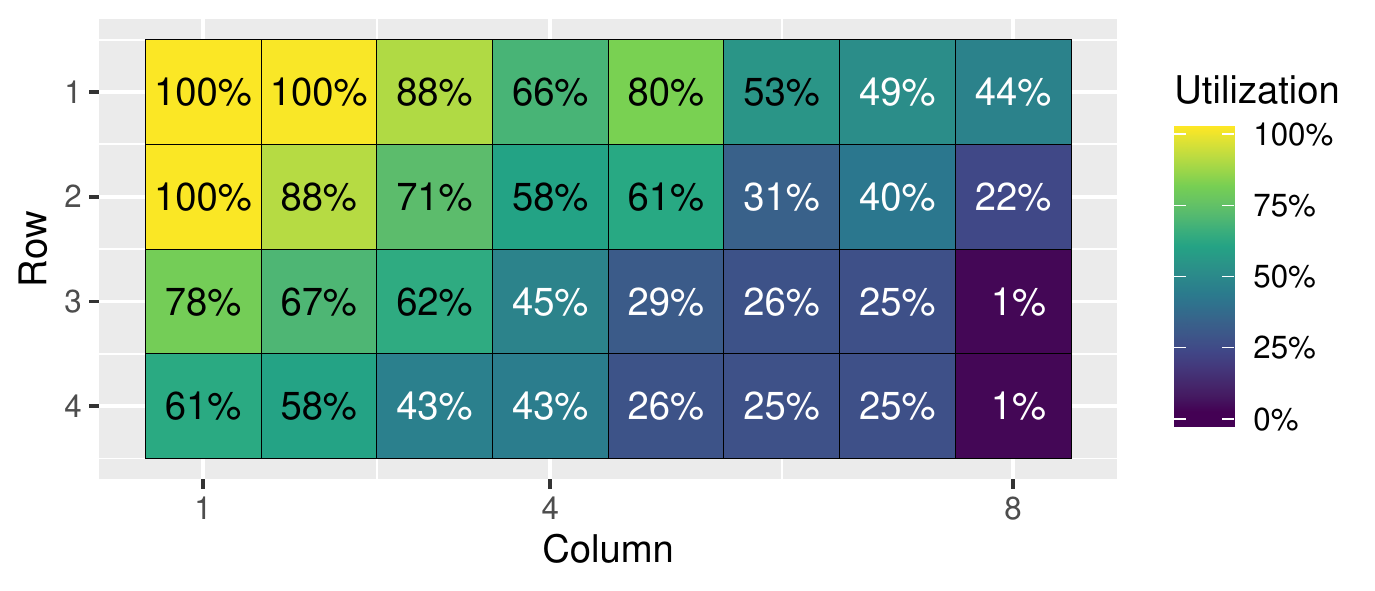}
    \vspace{-1.0\baselineskip}
    \caption{Utilization rate of the \gls{FU} in a 1D (only left-to-right data propagation) 4x8 \gls{CGRA} design when employing traditional mapping methods and executing a set of embedded applications.}
    \vspace{-1.0\baselineskip}
    \label{fig:intro_motivation}
\end{figure}

\textbf{Motivational Analysis:}
In \glspl{CGRA}, when employing traditional energy-efficient application mapping strategies \cite{Chen2014b, Dave2018, Watkins2016}, the allocation of the operations is typically biased towards one of the corners of the array.
The reason for that is, the allocation algorithms greedily select the first available \gls{FU} where an operation may execute in order to minimize the execution time, for instance, through reduced communication time.
Fig. \ref{fig:intro_motivation} from our experiments illustrates this phenomena by showing the average utilization of the \glspl{FU} in a rectangular \gls{CGRA} fabric (such as in \cite{Beck2014, Liu2015, Brandalero2019}) when executing a set of embedded benchmark applications.
As can be seen, the top-left-most \gls{FU} is used by 100\% of the \gls{CGRA} configurations, while the bottom-right-most one is used by only 1\% of the total configurations.
As a consequence of their high utilization, \glspl{FU} around the top-left corner undergo more stress phases over time and can age up to $10\times$ faster, 
leading to early-stage \gls{FU} failures that limit the \gls{ILP} exploitation and \gls{CGRA} performance.

\textbf{Proposed Approach:}
Ideally, the utilization should be uniformly distributed across the \gls{CGRA}'s \glspl{FU} to ensure a uniform aging rate and an extended system lifetime.
The \glspl{FU} with low utilization thus represent a \emph{utilization budget} that can be leveraged to slow down aging in the most stressed components.
Towards this goal, we propose a novel configuration allocation procedure that supports the automatic run-time movement of \gls{CGRA} configurations through the reconfigurable fabric, improving over traditional aging-unaware allocation strategies.
The approach is implemented by deploying important, yet low-cost extensions to the \gls{CGRA}'s reconfiguration logic and datapath that allow configuration movement without any significant performance overhead, effectively distributing the \gls{FU}'s utilization more uniformly across the fabric.

\textbf{In summary, this work makes the following contributions}:
\begin{itemize}[leftmargin=*]
    \item We propose a novel utilization-aware configurations allocation strategy for \glspl{CGRA} that automatically slows down the \gls{NBTI} aging effects by balancing the utilization of the \glspl{FU}  (Section III.A).
    \item We introduce the required architectural extensions to support the proposed allocation approach in a state-of-art \gls{CGRA} \cite{Brandalero2019} (Section III.B);
    \item We show how our strategy can increase the \gls{CGRA}'s lifetime by $2.4\times$ even under resource-constrained scenarios while introducing negligible performance and less than 10\% area overhead (Section IV).
\end{itemize}
\section{Background and Related Work}
\label{sec:related}

\subsection{Hardware Aging}

\gls{NBTI} is one of the fundamental aging phenomena affecting PMOS transistors.
Setting $V_{gs}$ to $-V_{dd}$ (i.e. switching the transistor on) leads to an increase in the \gls{Vt} which is commonly referred to as \emph{short-term aging}.
When the stress is released by setting $V_{gs}$ to $0$, the increase in \gls{Vt} is only \emph{partially} recovered, thereby leading to a continuous delay degradation over a long period of time (i.e., \emph{long-term aging}) \cite{Gnad2015, Alam2005, Bernstein2006}.
Recent evaluations suggest nearly 10\% increase in circuit delay after 3 years \cite{Oboril2012} and 20\% increase after 10 years \cite{Tiwari2008}, or complete wear-out in less than 3 years even for very low stress rates \cite{Paterna2013}.

The default strategy to prevent premature system failure is to ship designs with timing guardbands, setting a nominal frequency lower than the maximum one and accounting for the wear-out effects over several years \cite{Rahimi2013, Lefurgy2013}.
Architectural-level aging-mitigation strategies usually employ utilization balancing of the processing resources.
For that, however, a regular structure is required, limiting the application of the approach to multi-core \cite{Gnad2015, Paterna2013, Muck2017} and \gls{GPU} \cite{Rahimi2013a, Chen2014a, Lee2017} architectures.

In this work, we use a predictive model for \gls{NBTI} aging based on a prominent related work \cite{Henkel2013} that gives the long-term \gls{NBTI}-induced \gls{Vt} as a function of the \gls{Vdd}, Temperature ($T$), time ($t$) and duty cycle ($d$, which is equivalent to the utilization rate $u$ of an \gls{FU}) -- see Eq. \ref{eq:long_term_aging}.
The increase in delay can then be approximated to first order as the relative increase in \gls{Vt}. 

\begin{equation}
\begin{aligned}
\Delta{V_t} = %
0.005 \times %
e^{-1500/T} \times V_{dd}^4 \times t^{1/6} \times u^{1/6}
\end{aligned}
\label{eq:long_term_aging}
\end{equation}

\subsection{\glsentrylongpl{CGRA}}

Reconfigurable architectures, in particular, \glspl{CGRA}, represent an attractive solution for energy-efficient execution even for single-threaded applications since they enable hardware customization at run time to match different computational requirements \cite{Compton2002, Wijtvliet2016}.
Application code can be mapped to \glspl{CGRA} either \emph{statically} \cite{Hamzeh2012, Dave2018} (at compile time) or \emph{dynamically} \cite{Clark2004, Beck2014, Liu2015, Watkins2016} (at run time).
Dynamic mapping approaches present several advantages.
First, they enable the automatic acceleration of binaries after deployment, without the need to recompile.
Second, they can leverage dynamic information (such as how often each application region is executed) for optimizing the configurations towards the energy-efficiency target.

\begin{figure}
    \centering
    \includegraphics[width = 1\linewidth]{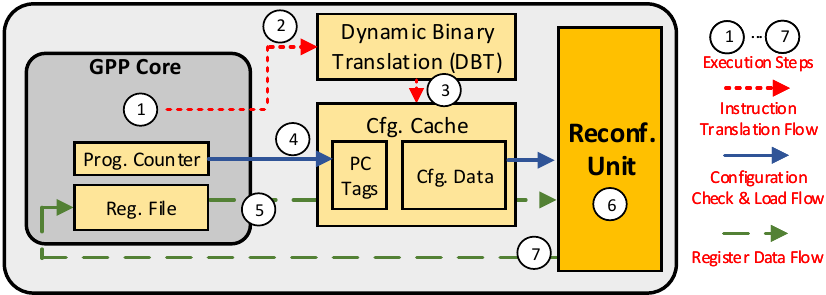}
    \vspace{-1.5\baselineskip}
    \caption{Overview of the \gls{TransRec} system \cite{Brandalero2019}, which serves as a baseline for the implementation of the utilization-aware allocation proposed in this work.}
    \label{fig:drawing_system}
    \vspace{-1\baselineskip}
\end{figure}

An example of such a system supporting dynamic mapping is the \gls{TransRec} system \cite{Brandalero2019}, which we use as a baseline for the proposed aging-aware load balancing strategy (described next in Section III).
\gls{TransRec} consists of a \gls{GPP} core, a tightly-coupled \gls{CGRA}-based reconfigurable unit, and a hardware-implemented \gls{DBT} module that automatically transforms binary code at run time into configurations for \gls{CGRA} execution.
An overview of this system and its execution process is presented in Fig. \ref{fig:drawing_system} and detailed next.
An application begins its execution on the \gls{GPP} core (Step 1).
As instructions finish their execution, they are sent to the \gls{DBT} module (Step 2), which interprets their semantics, finds the dependencies among them, and allocates them into a \gls{CGRA} configuration. A CGRA configuration is composed of a sequence of instructions that are then saved in a dedicated configuration cache and indexed by the PC of the first instruction of that sequence, for posterior acceleration (Step 3).
Therefore, while the \gls{GPP} executes the application, the DBT also continually checks the configuration cache for a matching configuration for the next instruction sequence using the \gls{PC} (Step 4).
When entry one is found, it is offloaded from the configuration cache along with the input register values coming from the \gls{GPP} (Step 5) and executed in the \gls{CGRA} (Step 6)
After the execution is completed, the output registers' results are written back and committed \emph{in program order} to the \gls{GPP} (Step 7).
This execution model enables on-the-fly acceleration without changing the application binary.

We present the detailed architecture of TransRec's reconfigurable unit in Section III, along with the required extensions to support the proposed aging-mitigation approach.

\subsection{Research Opportunities}

There is only a limited body of work on aging mitigation in \glspl{CGRA}.
Previous work has modified a static mapping strategy for aging-aware allocation \cite{JiangyuanGu2017}, or used aggressive voltage underscaling to achieve low energy consumption while also reducing aging \cite{Afzali-Kusha2018}.
This static approach, however, is limited to new applications with availability of source code, requires redeployment, and is unaware of dynamic input-dependent information that affects the execution.
Another work uses voltage aggressive underscaling to achieve low energy consumption and also reduce aging, given the dependence of \gls{NBTI} on the supply voltage \cite{Afzali-Kusha2018}.
The approach proposed in this work is complementary to \cite{Afzali-Kusha2018} since \gls{NBTI} is addressed by balancing the utilization of the processing resources and therefore increasing the stress-to-recovery ratio, which affects aging as per Eq. \ref{eq:long_term_aging}.
\section{Proposed Utilization-Aware Allocation Strategy}
\label{sec:system}

As described in Section II, current application-to-\gls{CGRA} mapping approaches are limited in their ability to address \gls{NBTI} aging.
In an ideal scenario, an aging-aware strategy should not allocate the application's instructions onto the lowest-health \glspl{FU} in the \gls{CGRA} fabric.
However, detecting the optimal allocation at run time may turn out to be prohibitively expensive; while supporting random allocations (and therefore achieve a uniform distribution over time) in the \gls{CGRA} fabric with a complex interconnection network may severely impact performance.

\begin{figure}[t]
    \centering
    \includegraphics[width=0.9\linewidth]{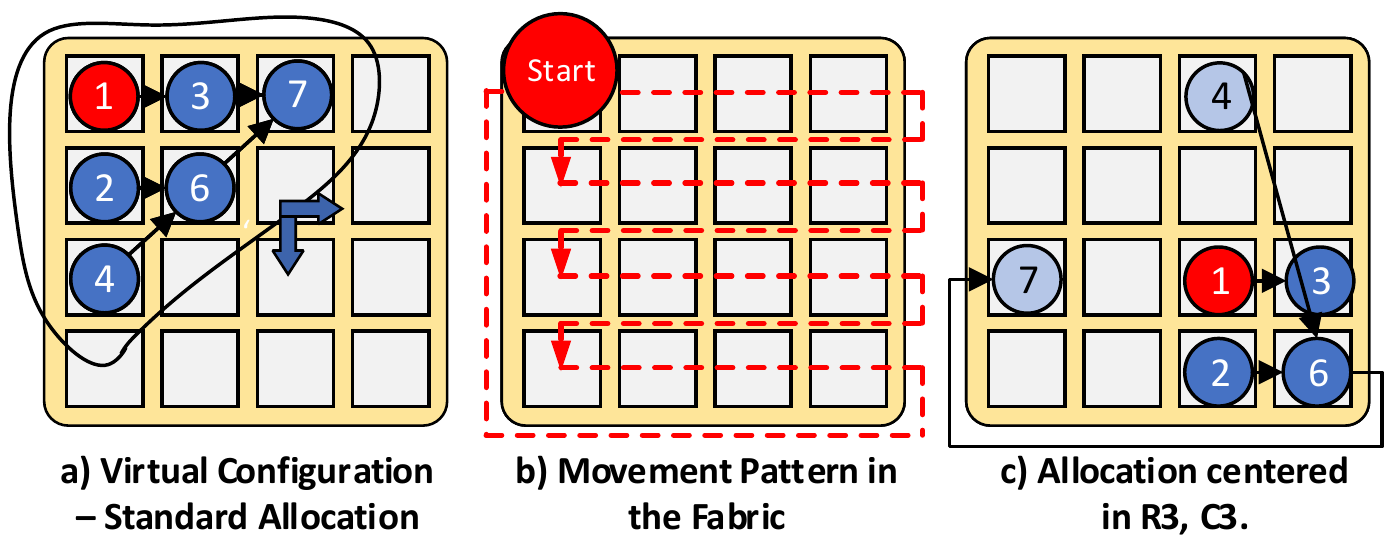}
    \caption{Overview of the two-step \emph{configuration generation} and \emph{configuration allocation} approach.}
    \vspace{-1\baselineskip}
    \label{fig:drawing_2steps}
\end{figure}

We propose a lightweight yet effective alternative: a utilization-aware configuration allocation strategy that enables automatic aging mitigation.
The strategy is implemented directly in the hardware, and therefore requires no changes to the precompiled \gls{CGRA} binaries or dynamic mapping strategies.
An overview is provided in Fig. \ref{fig:drawing_2steps}.
Given a previously generated \gls{CGRA} configuration (using static or dynamic methods), which we refer to as a \emph{virtual configuration} (See Fig. \ref{fig:drawing_2steps}a), the allocation approach consists of moving this configuration through the fabric horizontally and vertically whenever a new execution takes place.
To do so, we move the position of the configuration \emph{pivot} (red circle in Fig. \ref{fig:drawing_2steps}) for each new execution following the pattern depicted in Fig. \ref{fig:drawing_2steps}, which covers all of the reconfigurable fabric.
By moving the \emph{pivot}, the entire configuration moves with it, and a more uniform utilization of the entire fabric should be achieved.
To cover the entire \gls{CGRA}, \emph{wrap-around} as shown in Fig. \ref{fig:drawing_2steps}c is also supported.

We proceed with a description of the \gls{TransRec} \gls{CGRA}, which serves as a use-case on top of which we implement the proposed approach.
After that, we show the required architectural extensions to support the proposed approach.

\subsection{Baseline Implementation}

The \gls{TransRec} \gls{CGRA} is a matrix of \glspl{FU} composed entirely of combinational logic and divided into rows and columns -- see Fig. \ref{fig:cgra_detail}.
Data propagates from left to right, so that each \gls{FU} occupies a row and a sequence of columns (according to its latency). 
Since the design is highly regular, it is easily configurable for exploiting distinct ranges of \gls{ILP}.
For the technology under consideration in this work, \glspl{ALU} have the latency of half a processor cycle due to their simplicity and correspond to a single column.
Loads and stores are constrained by the data cache, with one read and one write, and take two processor cycles to complete (or, equivalently, four columns). As shown in the figure, while a single load operation or store is executed, two chains of four data-dependent \gls{ALU} operations may be executed concurrently.

\glspl{FU} communicate via context lines, initially fed by values from the input context.
Before each \gls{FU}, a crossbar selects the context line that feeds the inputs.
After each \gls{FU}, another crossbar selects, for each context line, which values from the FUs of the current column are propagated to the next ones.

\subsection{Microarchitectural Extensions}

To implement the instruction rotation mechanism with no performance overheads, we need to modify the interconnection network to allow:
\begin{itemize}
    \item the allocation of a previously-generated configuration into the \gls{CGRA} fabric to start at an arbitrary column and row;
    \item the allocation to ``\emph{wrap-around}'', so that operations in the last row and column can propagate their results to the first row and column (similarly to how circular buffers are designed).
\end{itemize}

Moreover, the implementation of the approach requires changes both in the reconfiguration logic and in the datapath.
Two operations must be supported: \emph{vertical movement} and \emph{horizontal movement}. 

For the \emph{vertical movement} operation, we need to only change the \gls{CGRA}'s reconfiguration logic.
This is the structure shown in Fig. \ref{fig:drawing_reconflogic}a, which is responsible for receiving the configuration bits from the configuration cache and forwarding them to the appropriate columns and rows.
It works as follows.
Input to the reconfiguration logic are $n$ \emph{configuration lines}, each of which contains the reconfiguration bits for an entire column.
Columns continuously monitor the \emph{configuration lines} such that column $i$ is connected to line $i \ mod \ n$.
For $n = 4$, for instance (as in Fig. \ref{fig:drawing_reconflogic}), columns $0, 4, 8 \ldots$ are connected to bus $0$, columns $1, 5, 9, \ldots$ to bus $1$ and so on.
A \emph{control unit} sends \emph{write} signals so that the context registers for columns $i, \ldots, i+n$ are reconfigured in cycle $i$.
In the example, columns $1 \ldots 4$ would all be reconfigured in the first cycle.

In the implementation supporting the configuration movement approach, the reconfiguration logic is extended with multiplexers to allow each column $i$ to fetch the values in any of the $n$ configuration lines.
Fig. \ref{fig:drawing_reconflogic}b shows this new multiplexer (in purple), which allows the column to receive the configuration bits for any of the configuration lines, not only the first one (shown in red).
This microarchitectural change enables the horizontal movement of the configurations through the fabric.

Lastly, Fig. \ref{fig:drawing_reconflogic}c shows (in purple) the required extensions necessary to support the vertical movement of the configuration.
Considering that the input configuration bits are responsible for an entire column, which includes the input multiplexers, the \glspl{FU} and the output multiplexers, we extend the microarchitecture with barrel shifters for each of these structures (shown in purple).
Although barrel shifters are known to introduce significant latency overheads, the number of rows in such a \gls{CGRA} is usually small ($\leq 8$) due to the limited application \gls{ILP}, so such overhead is only marginal, as our results will show.

\begin{figure}[t]
    \centering
    \includegraphics[width=0.9\linewidth]{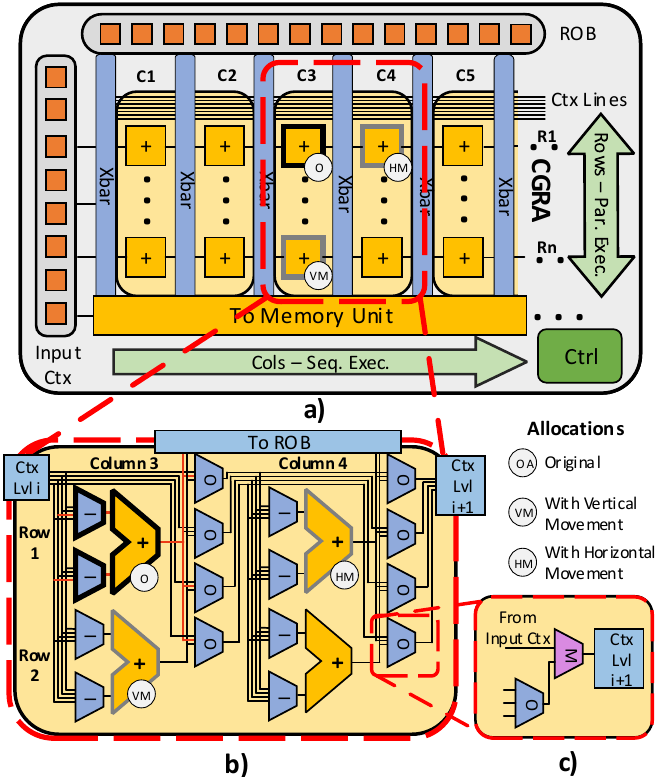}
    \caption{Detailed view of (a) the \gls{CGRA}-based Reconfigurable Unit, (b) two columns with reconfigurable datapath and the option to move the original \gls{FU} allocation (O) vertically (VM) and horizontally (HM), and (c) the required addition to the datapath to support horizontal movement (shown with R).}
    \label{fig:cgra_detail}
    \vspace{-1\baselineskip}
\end{figure}

\begin{figure}[t]
    \centering
    \includegraphics[width=0.8\linewidth]{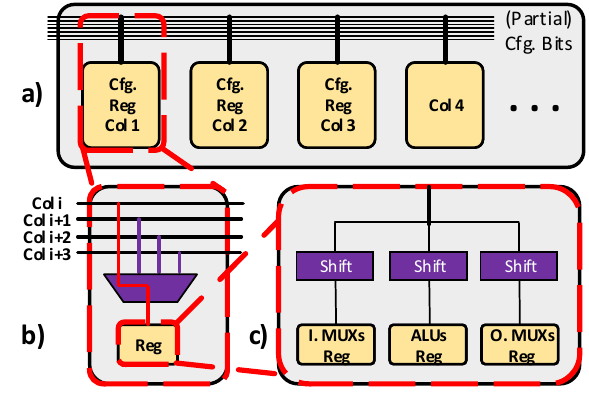}
    \caption{Detailed view of the reconfiguration logic, with (a) the configuration lines feeding all columns, (b) a column listening to a particular configuration line, and (c) the configuration bits feeding to the input multiplexers, \glspl{FU} and output multiplexers.}
    \label{fig:drawing_reconflogic}
    \vspace{-1\baselineskip}
\end{figure}

The proposed approach, in particular, the horizontal movement operation, also requires the introduction of one new set of multiplexers for each column and the context line in the datapath.
An additional 2:1 multiplexer (shown in purple in Fig. \ref{fig:cgra_detail}) selects between the previous line and the initial input context.
This multiplexer is deployed to enable the wrap-around operation, i.e., so that a configuration may start the execution from an arbitrary column.
The multiplexer from the last column then feeds back into the first one to enable the wrap-around of the operations (i.e., operations in the last column can communicate to the first column).

\section{Evaluation Methodology}
\label{sec:methodology}

We evaluate the proposed utilzation-aware allocation for its lifetime-extension capabilities and its performance and area overheads, comparing these results to the utilization-unaware implementation.
We target the RISC-V \gls{ISA} due to the availability of open \gls{HDL} processor designs that allow for high-accuracy area and power estimations.

\subsection{Tools and Design Flow}

For performance evaluation and fast design-space exploration, we extended the gem5 cycle-accurate simulator \cite{Binkert2011} with the TransRec implementation \gls{BT} and \gls{CGRA} designs based on \cite{Brandalero2019}, using the \emph{TimingSimple} CPU to model a single-issue core.
10 benchmarks\footnote{The following subset of \emph{mibench} was selected due compatibility with the simulation toolflow: \emph{bitcount}, \emph{CRC32}, \emph{dijkstra}, \emph{qsort}, \emph{rijndael-e}, \emph{sha}, \emph{stringsearch}, \emph{susan} (corners, edges and smoothing).} from mibench \cite{Guthaus2001}, typically found in the embedded domain, compiled for the RISC-V \gls{ISA} with \textsc{-O3} and running the ``small input set'' were used in the evaluation.

For area and energy evaluation, an \gls{HDL} prototype of the system built on top of the Rocket core \cite{Asanovic2016} was developed and synthesized with Cadence RTL Compiler and NanGate's 15nm standard cell library \cite{Martins2015}.
Results for the caches were estimated using FinCACTI \cite{Shafaei2014}.

Finally, to evaluate the aging, the model described earlier in Section II (Eq. \ref{eq:long_term_aging}) was used {\toreview to estimate the impact of utilization on the NBTI-induced increase in $V_t$.
The $V_t$ degradation causes a linear increase in delay, according to the model of \cite{Henkel2013}}. 
We consider that the product's end-of-life is determined by the aging rate of the most stress-induced component in the design (i.e., the \gls{FU} with the highest utilization).
A worst-case delay degradation of 10\% over 3 years was considered as estimated in the literature \cite{Tiwari2008, Rehman2014}.

\subsection{Design Space Exploration}

\begin{figure}
    \centering
    \includegraphics[width=1\linewidth]{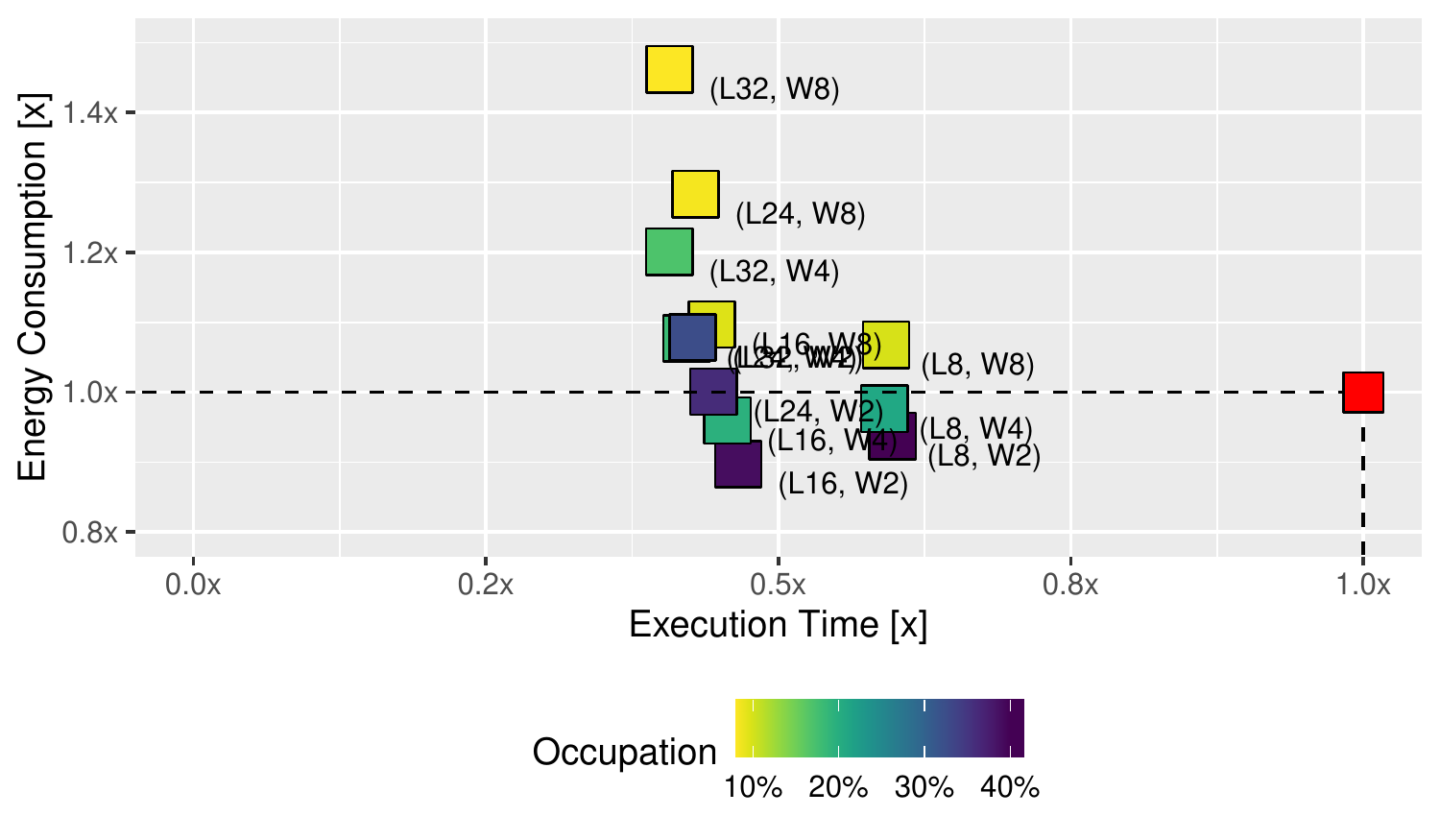}
    \vspace{-2\baselineskip}
    \caption{System-wide performance, energy consumption and average occupancy for different sizes of \gls{CGRA} fabric.}
    \vspace{-1\baselineskip}
    \label{fig:design_space}
\end{figure}

An initial exploration of the unmodified \gls{TransRec} architecture design space was carried out to select interesting design points to evaluate the aging effects.
This exploration covered different sizes of the \gls{CGRA} fabric (length as number of columns $L$, representing sequential execution, varied from 8 to 32, and width as number of rows $W$, representing parallel execution, varied from 2 to 8) and its impact onto the execution time, the energy consumption and the average utilization of the \glspl{FU} (all compared to a stand-alone Rocket \gls{GPP} core).
Fig. \ref{fig:design_space} shows the result of this exploration.
The red square with $1\times$ represents the performance and energy consumption of the \gls{GPP} alone, and the others the \gls{TransRec} system.
From this analysis we select the following designs of interest:
\begin{itemize}
    \item $(L16, W2)$, with $2.14\times$ speedup, $10\%$ reduction in energy consumption and an average utilization of $39.7\%$. We name this scenario $BE$ (\emph{best energy consumption}).
    \item $(L32, W4)$, with $2.45\times$ speedup, $20\%$ increase in energy consumption and an average utilization of $17.8\%$. We name this scenario $BP$ (\emph{best performance}).
    \item $(L32, W8)$, with $2.45\times$ speedup (same as above), $46\%$ increase in energy consumption and an average utilization of $8.9\%$. We name this scenario $BU$ (\emph{best (lowest) utilization}).
\end{itemize}

\section{Results and Discussion}
\label{sec:results}

\subsection{Utilization and Lifetime Extension}

\begin{figure}
    \centering
    \includegraphics[width=0.95\linewidth]{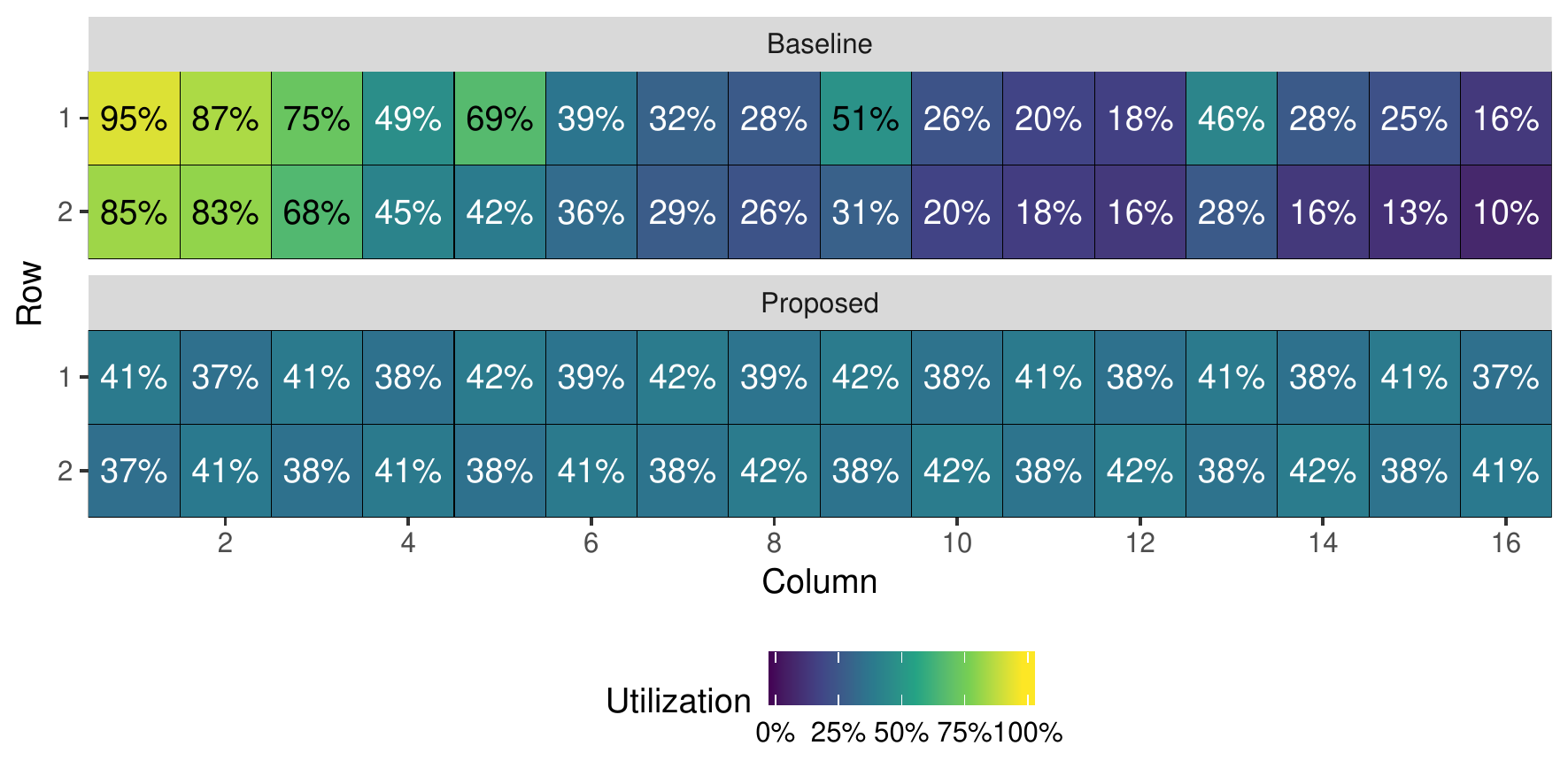}
    \vspace{-1\baselineskip}
    \caption{Average utilization of each \gls{FU} in the traditional and proposed approaches (left side) along with probability density plot for the utilization rate. Results for the BE scenario (16x2 design).}
    \vspace{-1\baselineskip}
    \label{fig:plot_scenario_BE}
\end{figure}

Fig. \ref{fig:plot_scenario_BE} shows the utilization of the \gls{CGRA} fabric in the \emph{BE} scenario, for both the baseline allocation strategy (on the top) as well as the utilization-aware allocation (on the bottom).
As initially noted in Section I, the lower columns and rows present a higher utilization than the upper ones when employing traditional allocation strategies.
With the proposed utilization-aware allocation, the utilization is more evenly distributed across the fabric, avoiding high stresses in a single component.
The maximum utilization drops from 94.5\% in the baseline to 41.2\% in the proposed approach.

\begin{figure}
    \centering
    \includegraphics[width=0.95\linewidth]{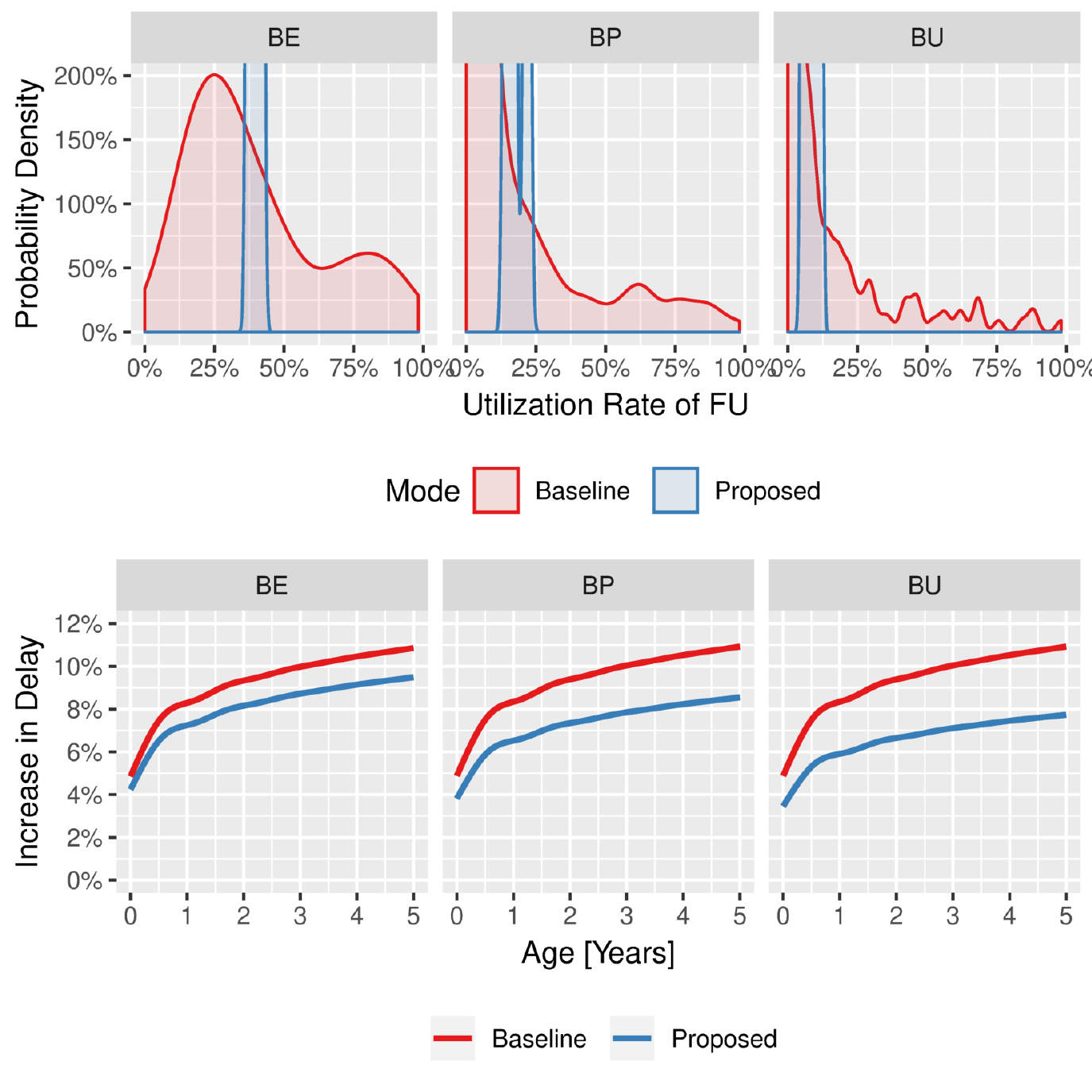}
    \vspace{-1\baselineskip}
    \caption{\gls{NBTI}-induced increase in delay over the years for different \gls{CGRA} sizes and considering the utilization-unware allocation and the proposed one.}
    \vspace{-1\baselineskip}
    \label{fig:utilization_aging_curves}
\end{figure}

Due to space limitations, rather than presenting the same figure for the \emph{BP} and \emph{BU} scenarios (which represent larger \gls{CGRA} fabrics), we show in the upper part of Fig. \ref{fig:utilization_aging_curves} the \emph{probability density function} for the utilization of a \gls{FU} in the three scenarios, for both the baseline as well as the proposed allocation approach.
These plots show that larger fabrics such as in the \emph{BU} scenario have a significant amount of \gls{FU} with small utilization when using the baseline allocation, increasing the potential of the proposed approach.


Given these distributions and the highest utilization for each scenario (which ultimately determines the first component to fail), we use Eq. \ref{eq:long_term_aging} to estimate the improvements in the aging rate of the design.
The lower part of Fig. \ref{fig:utilization_aging_curves} shows the delay increase over time due to \gls{NBTI} aging for the baseline and proposed approaches.
Larger designs present the highest difference, as they offer a higher \emph{utilization budget} for balancing.
For the smallest design (\emph{BE}), the aging rate is slowed down by $2.29\times$, and the system presents a performance degradation of 10\% only in 7 years rather than in 3.
Larger designs lead to even better improvements in the product's lifetime as they offer even more resources for utilization balancing.
Table I summarizes these findings.

\begin{table}[]
\vspace{-1\baselineskip}
\caption{Utilization and lifetime improvements for each scenario.}
\label{tab:utilization_aging_scenarios}
\vspace{-1\baselineskip}
\centering
\begin{tabular}{@{}ccccc@{}}
\toprule
Scenario & Avg. Util & \begin{tabular}[c]{@{}c@{}}Baseline\\ Worst Util.\end{tabular} & \begin{tabular}[c]{@{}c@{}}Proposed\\ Worst Util.\end{tabular} & Lifetime Improv.          \\ \midrule
BE       & 39.7\%    & 94.5\%                                                         & 41.1\%                                                         & $2.29\times$ \\
BP       & 17.10\%   & 98.1\%                                                         & 22.4\%                                                         & $4.37\times$ \\
BU       & 8.5\%     & 98.1\%                                                         & 12.3\%                                                         & $7.97\times$ \\ \bottomrule
\end{tabular}
\end{table}

\subsection{Area Overhead}

As described in Section III, the proposed approach requires extending the reconfiguration logic and the datapath.
Table II presents the area results for the implementation of the \emph{BE} scenario with and without the architectural extensions.
The area overhead was found to be below 10\%.
Moreover, considering only a single column in the design, both the baseline and the proposed version were able to reach the same minimum latency of 120ps (for one column, as shown in the left side of Fig. \ref{fig:cgra_detail}b).
These results suggest that the introduced modifications do not affect the maximum frequency of the design, which is likely to run with a frequency below the maximum one due to power consumption constraints.  

\begin{table}[]
\centering
\caption{\gls{CGRA} area overhead (BE Scenario).}
\vspace{-1\baselineskip}
\label{tab:my-table}
\begin{tabular}{@{}llc@{}}
\toprule
                                           & \multirow{2}{*}{Baseline} & \multirow{2}{*}{Modified} \\
                                           &                           &                           \\ \midrule
\multicolumn{1}{c}{\textbf{Area{[}$\mu^2${]}}} & \multicolumn{1}{c}{$28,995$} & $30,199 \ (+4.15\%) $           \\
\multicolumn{1}{c}{\textbf{\# Cells}}      & \multicolumn{1}{c}{$79,540$} & $83,083 \ (+4.45\%)$            \\ \bottomrule
\end{tabular}
\vspace{-2\baselineskip}
\end{table}

\section{Conclusions and Future Work}
\label{sec:conclusions}

In this work, we have proposed a utilization-aware allocation strategy for aging mitigation in \glspl{CGRA}.
The strategy is based on rotating the \emph{virtual configurations}, which were originally targeted for performance/energy efficiency, throughout the fabric to balance the stress-to-recovery rates of the individual \glspl{FU}, and thereby reduce the aging rate of the whole design.
{
\toreview 
To enable this, the configuration bits are shifted at configuration load time by using the structures shown in Fig. 5.
}
By doing so, our approach increases the lifetime of the design by $2.29\times$--$7.97\times$ for different design sizes. As a future work, we will implement the improved rotation techniques and use run-time aging information to adapt the allocation strategy dynamically. We will also evaluate homogeneous and heterogeneous multi-core scenarios.

\section*{Acknowledgements}

{
\toreview
This work is supported in part by the German Research Foundation (DFG) as part of the priority program "Dependable Embedded Systems" (SPP1500 - http://spp1500.itec.kit.edu), and in part by the Postoc Network Brandenburg (https://www.postdoc-network-brandenburg.de/en/).
}

\bibliographystyle{IEEEtran}
\bibliography{99-References}

\end{document}